\documentclass[reprint,prl,longbibliography,lengthcheck]{revtex4-2} 

\usepackage{graphicx}
\usepackage{amssymb}
\usepackage{amsfonts}
\usepackage{amsmath}
\usepackage{amsbsy}
\usepackage{mathrsfs} 
\usepackage{verbatim} 
\usepackage{bm}
\usepackage{color}
\usepackage[colorlinks=true,citecolor=blue]{hyperref}
\usepackage{lineno}

\usepackage{mymacros}

\begin{document} 

\title{Universal square-root-log scaling for slender-body end effects}

\author{Gunnar Peng$^1$ and Ory Schnitzer$^2$}
\affiliation{$^1$Department of Mathematics, University College London, London WC1E 6BT, UK}
\affiliation{$^2$Department of Mathematics, Imperial College London, 180 Queen's Gate, London SW7 2AZ, UK}

\begin{abstract}
The charge on a conducting cylinder becomes uniformly distributed at large aspect ratios, albeit very slowly and with nonuniformity persisting near the ends. We show that the charge density and field magnitude are enhanced near the ends by a factor scaling as the square-root of the logarithm of the aspect ratio. This scaling is obtained by locally resumming a perturbation-series solution to the integral equation of slender-body theory. The same scaling applies to a broad class of ``truncated'' slender bodies---including cylinders and shapes tapered near the ends on the cross-sectional scale---as well as other physical setups. We validate this scaling through boundary-element simulations for cylinders with flat and hemispherical caps, and demonstrate its applicability to diffusion, plasmonics and Stokes flow.
\end{abstract}

\maketitle

In one of his final papers \citep{Maxwell:77}, Maxwell considered the capacitance and charge distribution of thin conductors in an unbounded dielectric. He postulated that the linear charge density along a wire ``without breadth'' is uniform, finding that for a finite cylinder this limit is approached logarithmically slowly, with the deviation from uniformity becoming ``more confined to the parts near the ends.''  While Maxwell's problem has been frequently revisited \citep{Smythe:56,Miles:67,Wang:84,Djordjevic:85,Griffiths:96,Jackson:00,Bonfim:01,Jackson:02,Van:03,Fikioris:17,Lekner:21,Sousa:26}---most successfully within the paradigm of slender-body theory \citep{Handelsman:67,Hinch:book,Keller:03charge,Sellier:99,Chadwick:10,Tuck:12}---a systematic asymptotic description of charge crowding near the ends has remained elusive for cylindrical, as opposed to spheroidal, geometries. Here, we resolve this long-standing gap, unraveling a surprisingly intricate asymptotic structure relevant to broad classes of slender geometries and physical systems. 

\begin{figure*}
\begin{center}
\includegraphics[scale=0.55,trim={1.2cm 0 0 0}]{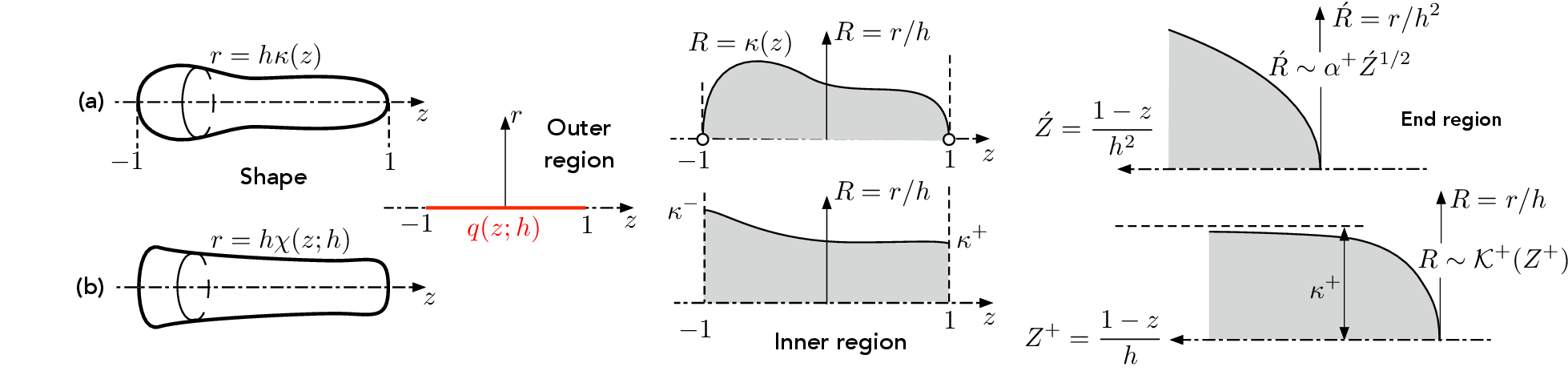}
\caption{Slender-body theory for (a) needle-like shapes tapered on the longitudinal scale, including prolate spheroids; and (b) truncated shapes tapered on the cross-sectional scale, including flat- and round-ended cylinders.}
\label{fig:sketch}
\end{center}
\end{figure*}

We begin by outlining the application of slender-body theory to the electrostatics of a generic thin conductor. Consider a body of revolution of half-length $L$, situated within an unbounded dielectric of permittivity $\epsilon$ and held at a fixed potential $V$ relative to infinity. This axisymmetric problem requires solving Laplace's equation for the exterior potential $\varphi$ (normalized by the voltage $V$), subject to $\varphi=1$ on the conductor surface and far-field decay. We employ cylindrical coordinates $(Lr,\phi,Lz)$, with the interval $z\in(-1,1)$ along the symmetry axis coinciding with the centerline. We first recapitulate the case where the body is ``needle''-like: smoothly tapered, with its cross-sectional radius varying strictly on the longitudinal $O(1)$ scale (e.g., a prolate spheroid). The boundary can then be defined as 
\begin{equation}\label{kappa exact}
r=h\kappa(z),
\end{equation}
where the thickness profile $\kappa(z)$ is independent of the slenderness parameter $h$ and satisfies $\kappa(z)\sim \alpha^{\pm}\sqrt{1\mp z}$ as $z\to\pm1$, respectively, where $\alpha^\pm$ are constants [Fig.~\ref{fig:sketch}(a)]. 

In the slender-body limit $h\ll1$, the body collapses onto the centerline [Fig.~\ref{fig:sketch}]. This suggests an approximation $\varphi\sim \tilde{\varphi}$ formed by continuously distributing monopole sources along it, 
\begin{equation}\label{outer}
\tilde{\varphi} = \frac{1}{4\pi}\int_{-1}^1\frac{q(\xi)}{\sqrt{r^2+(\xi-z)^2}}\,d\xi, 
\end{equation}
with $q(z)$ the apparent linear charge density (normalized by $\epsilon V$). This approximation satisfies far-field decay, but diverges logarithmically upon approaching the centerline,
\begin{equation}\label{local}
\tilde{\varphi} \sim \frac{q(z)}{2\pi}\ln\frac{2\sqrt{1-z^2}}{r}+\frac{1}{2\pi}\mathcal{S}[q](z) \quad \text{as} \quad r\to0,
\end{equation}
for $|z|<1$, with an $O(r^2\ln r)$ error and the $\mathcal{S}$-transform $\mathcal{S}[q](z)$ defined by the principal-value integral \citep{Blake:10}
\begin{equation}\label{S transform}
\mathcal{S}[q](z)=\frac{1}{2}\int_{-1}^1\frac{q(\xi)-q(z)}{|\xi-z|}\,d\xi.
\end{equation}
Given this divergence, Maxwell's idealization of a wire ``without breadth'' is actually ill-posed. 

To satisfy the equipotential boundary condition, the above ``outer'' approximation must be matched with an ``inner'' approximation holding near the body [Fig.~\ref{fig:sketch}(a)]. The latter region is analyzed by defining $\varphi(r,z)=\Phi(R,z)$ and taking the limit $h\to0$ while holding $z$ and the rescaled radial coordinate $R=r/h$ fixed (with $R>\kappa(z)$ and $|z|<1$). Under this radial rescaling, the boundary appears slowly varying, so that the Laplacian is dominated by its radial part. Thus, an inner approximation satisfying the boundary condition $\Phi(\kappa(z),z)=1$ is given by
\begin{equation}\label{inner}
\Phi \sim A(z)\ln\frac{R}{\kappa(z)}+1, 
\end{equation} 
where $A(z)$ is a function to be determined. 

Using \eqref{local}, the inner and outer approximations can be asymptotically matched to leading algebraic order in $h$, provided $q(z)$ and $A(z)$ are allowed to depend \emph{logarithmically} upon $h$; to reflect this, we henceforth denote these functions as $q(z;h)$ and $A(z;h)$. The matching then  shows that $A(z;h)=-q(z;h)/2\pi$, with $q(z;h)$ satisfying the  integral equation 
\begin{equation}\label{integral equation}
q(z;h)\ln\frac{2\sqrt{1-z^2}}{h\kappa(z)}+\mathcal{S}[q](z;h)=2\pi.
\end{equation}
The corresponding approximation for the capacitance (normalized by $\epsilon L$) is $\int_{-1}^1q(z;h)\,dz$. 

A prolate spheroid corresponds to the thickness profile $\kappa(z)=\sqrt{1-z^2}$. Since the $\mathcal{S}$-transform vanishes for constant functions, the solution to \eqref{integral equation} in this case is identically the uniform distribution $q(z;h)=2\pi/\ln(2/h)$, in agreement with the exact, precisely uniform, distribution for a spheroid \cite{Landau:Book:Electrodynamics}---up to an $O(h^2/\ln h)$ error. For other thickness profiles, the solution to \eqref{integral equation} can be obtained as a regular-perturbation expansion in inverse powers of the aspect-ratio logarithm $\mathcal{L}=\ln(1/h)\gg1$,
\begin{equation}\label{q expansion}
q(z;h) \sim \frac{1}{\mathcal{L}}\sum_{n=0}^{\infty} \frac{q_n(z)}{\mathcal{L}^n} \quad \text{as} \quad h\to0,
\end{equation}
where the first two terms are readily obtained as
\refstepcounter{equation}
\label{q0 and q1}
$$
q_0=2\pi, \quad q_1=-2\pi \ln \frac{2\sqrt{1-z^2}}{\kappa(z)},
\eqno{(\theequation{\mathrm{a},\mathrm{b}})}
$$
while higher-order terms can be calculated from the recurrence relation 
\begin{equation}\label{q recurrence} 
q_{n+1}(z)=-q_n(z)\ln\frac{2\sqrt{1-z^2}}{\kappa(z)}-\mathcal{S}[q_n](z).  
\end{equation}
Alternatively, \eqref{integral equation} can be solved numerically, thereby effectively summing the perturbation expansion. 

For the present class of needle-like shapes [see \eqref{kappa exact} \textit{et seq.}], the solution to \eqref{integral equation} is regular: $q(z;h)$ is smooth, with finite limits as $z\to\pm1$. Indeed, the logarithm in \eqref{integral equation} is regular under the assumed smoothness and square-root tapering of $\kappa(z)$, and the $\mathcal{S}$-transform can be shown to preserve regularity \footnote{A regular function on the interval $z\in[-1,1]$ can be expanded in Legendre polynomials, which are eigenfunctions of the $\mathcal{S}$-transform with eigenvalues $-H_n$, where $H_n$ is the $n$-th harmonic number ($H_0=0$). The fact that these eigenvalues grow only logarithmically as $n\to\infty$ implies that the $\mathcal{S}$-transform of a regular function is itself a regular function.}. 
Near the tips at $(r,z)=(0,\pm1)$, the inner and outer approximations both match with $O(h^2)$ end regions where the thickness scales as the axial distance from the respective end, and the boundary of the conductor is approximately a paraboloid [Fig.~\ref{fig:sketch}(a)]. This matching can be demonstrated analytically, with paraboloidal coordinates furnishing a closed-form approximation for the end-region potential to leading algebraic order \footnote{See \citep{Ruiz:19}. An alternative approach is to develop slender-body approximations that are uniformly valid including near the paraboloidal tips \citep{Tuck:64,Handelsman:67,Johnson:80}.}. 
We here simply observe that the regularity of the inner approximation implies that the linear charge density remains $O(\mathcal{L}^{-1})$ within the end regions; given the $O(h^2)$ scale of those regions, the surface charge density and electric field (normalized by $V/L$) there are $O(\mathcal{L}^{-1}h^{-2})$.
Because the end regions encompass an $O(h^4)$ surface area, their contribution to the net capacitance is $O(\mathcal{L}^{-1}h^2)$, thus subdominant to the $O(\mathcal{L}^{-1})$ body-scale contribution.

We now address Maxwell's problem of a thin conducting cylinder. In this case, the end regions are larger, corresponding to $r$ and $1\mp z$ being $O(h)$, rather than $O(h^2)$. This motivates a generalization from cylinders to a class of ``truncated'' slender shapes tapered on the $O(h)$ scale. Such geometries can no longer be expressed in the form \eqref{kappa exact}; instead, we define the boundary as 
\begin{equation}\label{chi def}
r=h\chi(z;h),
\end{equation}
where $\chi(z;h)$ is nonuniform as $h\to0$:
\begin{subequations}
\label{nonuniform shape}
\begin{gather}
\chi(z;h)\sim \kappa(z), \quad |z|<1\text{ fixed}, \\
\chi(z;h) \sim \mathcal{K}^{\pm}(Z^\pm), \quad Z^\pm=\frac{1\mp z}{h}>0\text{ fixed},
\end{gather}
\end{subequations}
with algebraically small errors [Fig.~\ref{fig:sketch}(b)]. Here $\kappa(z)$ is reinterpreted as a longitudinal-scale thickness profile having non-zero limits at the ends, say $\lim_{z\to\pm1}\kappa(z)=\kappa^{\pm}$; and $\mathcal{K}^{\pm}(Z^{\pm})$ are cap profiles vanishing at $Z^{\pm}=0$ and asymptotically matching with the longitudinal-scale profile: $\lim_{Z^{\pm}\to\infty}\mathcal{K}^{\pm}(Z^{\pm})=\kappa^{\pm}$.  For cylinders, $\kappa(z)\equiv 1$ and $\mathcal{K}^\pm(Z^\pm)=\mathcal{H}(Z^{\pm})$, where $\mathcal{H}$ denotes the Heaviside function; we shall refer to smooth bodies with $\kappa(z)\equiv1$ as ``round-ended cylinders.'' 

Both the outer  \eqref{outer} and inner \eqref{inner} approximations retain their form; consequently, so does the integral equation \eqref{integral equation} derived by matching them. Thus, to leading algebraic order in $h$, the asymptotic description of these regions remains unchanged, apart from the reinterpretation of the thickness profile $\kappa(z)$. As shown below, however, the regularity of the solution $q(z;h)$ to that integral equation is lost, modifying the matching with the end regions. 

To reveal the pathological form of this breakdown, we revisit the perturbation expansion for $q(z;h)$ in inverse-log powers [see \eqref{q expansion}--\eqref{q recurrence}]. The zeroth term $q_0$ is constant. With $\kappa(z)$ now having non-zero limits as $z\to\pm1$, the first term $q_1$ diverges logarithmically in those limits, $q_1\propto \ln(1\mp z)$ \footnote{Maxwell \citep{Maxwell:77} overlooked this local singularity; as discussed by \citet{Jackson:02}, Maxwell's approach relied on a variational approximation that projected the charge density onto a truncated basis of Legendre polynomials. This enforced artificial polynomial regularity at the ends, thereby obscuring the singular nature of the end behavior.}. At the next order, the recurrence relation \eqref{q recurrence} yields
\begin{equation}
q_2(z)=2\pi\left[\ln\frac{2\sqrt{1-z^2}}{\kappa(z)}\right]^2-\mathcal{S}[q_1](z).
\end{equation}
Directly evaluating $\mathcal{S}[q_1](z)$ shows that both terms on the right-hand side diverge $\propto [\ln(1\mp z)]^2$, as  does their sum. 

These observations suggest the hypothesis
\begin{equation}\label{qn ansatz}
q_n(z)\sim c_n^{\pm}\left[-\ln(1\mp z)\right]^n \quad \text{as} \quad z\to \pm 1,
\end{equation} 
which we confirm by induction. In the Supplementary Material (SM), we show that if \eqref{qn ansatz} holds for a given $n$, then 
\begin{equation}
\mathcal{S}[q_n](z)\sim -\frac{nc_n^{\pm}}{2(n+1)}[-\ln (1\mp z)]^{n+1} \quad \text{as} \quad z\to\pm 1.
\end{equation}
The recurrence relation \eqref{q recurrence} then confirms that \eqref{qn ansatz} is also satisfied for $q_{n+1}(z)$, with $c^{\pm}_{n+1}/c^{\pm}_n=(2n+1)/(2n+2)$. Since $c^{\pm}_0=2\pi$, we find that the coefficients at both ends are identical; denoting them as $c_n^{\pm}\equiv c_n$, we obtain
\begin{equation}\label{c sol}
c_n=\frac{2\pi}{4^n}\binom{2n}{n}.
\end{equation}

The result \eqref{qn ansatz} implies that near the ends \textit{all} terms in the inverse-log expansion \eqref{q expansion} become comparable---not just under the end-region scaling $1\mp z=O(h)$, where this is anticipated, but also under all intermediate algebraic scalings $1\mp z = O(h^\alpha)$, with $0<\alpha<1$. Across this intermediate regime, the \emph{leading-order} behavior is provided by the local resummation 
\begin{equation}\label{eta sum}
\bar{q}(\eta;h) \sim \frac{1}{\mathcal{L}}\sum_{n=0}^{\infty} c_n \eta^n \quad \text{as} \quad h\to0,
\end{equation}
where we have applied the change of variables $\bar{q}(\eta;h)=q(z;h)$ into the intermediate variable 
\begin{equation}\label{eta def}
\eta=\frac{1}{\mathcal{L}}\ln\frac{1}{1\mp z}.
\end{equation}
In \eqref{eta sum}, $\eta$ is held fixed within the interval $(0,1)$. The $O(1)$ longitudinal scale corresponds to $\eta=O(1/\mathcal{L})$, whereas the $O(h)$ end region corresponds to $1- \eta=O(1/\mathcal{L})$. 
Remarkably, \eqref{eta sum} can be explicitly summed, using \eqref{c sol}, yielding the leading-order intermediate behavior 
\begin{equation}\label{intermediate in eta}
\bar{q}(\eta;h) \sim \frac{2\pi}{\mathcal{L}}\frac{1}{\sqrt{1-\eta}} \quad \text{as} \quad h\to0.
\end{equation}
Written in terms of the end-region variables $Z^{\pm}=(1\mp z)/h$ [Fig.~\ref{fig:sketch}(b)], this intermediate behavior becomes
\begin{equation}\label{intermediate final}
q\sim \frac{2\pi}{\sqrt{\mathcal{L}\ln Z^{\pm}}}. 
\end{equation}
Notably,  \eqref{intermediate final} is independent of both the thickness profile $\kappa(z)$ and the cap profiles $\mathcal{K}^{\pm}(Z^{\pm})$. 

The behavior \eqref{intermediate final} implies the following key physical scalings in the end regions. The linear charge density is $O(\mathcal{L}^{-1/2})$, and thus the surface charge density and electric field are $O(\mathcal{L}^{-1/2}h^{-1})$. Integrating the end-region charge density yields an $O(\mathcal{L}^{-1/2}h)$ contribution to the net capacitance. While larger than the corresponding contribution for needle-like shapes, this remains subdominant. 

Although the end regions do not appreciably affect the net capacitance, their locally amplified fields govern threshold phenomena such as dielectric breakdown and field emission. We have seen that slender conductors generically exhibit a twofold lightning-rod effect. Relative to the characteristic $O(1)$ scaling formed by the imposed voltage and the body length, the inner electric field along the bulk of the conductor scales as $O(\mathcal{L}^{-1}h^{-1})$. Approaching the end regions, this field is further amplified, reaching $O(\mathcal{L}^{-1}h^{-2})$ for needle-like shapes and $O(\mathcal{L}^{-1/2}h^{-1})$ for truncated shapes. While this latter scaling seems comparatively mild, truncated shapes often exhibit a \emph{threefold} effect. Indeed, for a cylinder with flat faces, the well-known $3\pi/2$ corner singularity dictates that the field further diverges $\propto (Z^{\pm})^{-1/3}$ upon approaching the rim. If this corner is instead rounded with a radius of curvature $\varrho h\ll h$, the field saturates at $O(\mathcal{L}^{-1/2}h^{-1}{\varrho}^{-1/3})$. Consequently, the local field amplification becomes comparable between needle-like and truncated shapes for $\varrho$ of order $\mathcal{L}^{3/2}h^3$.

\begin{figure}[t!]
\begin{center}
\includegraphics[scale=1.1,trim={0.5cm 0.2cm 0cm 0}]{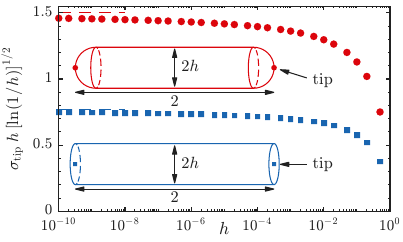}
\caption{Boundary-element simulations for cylinders with flat and hemispherical caps, with $\sigma_{\text{tip}}$ the tip surface charge density normalized by $\epsilon V /L$. The dashed lines correspond to $h=10^{-100}$.}
\label{fig:bem}
\end{center}
\end{figure}
We have validated our scaling theory through boundary-element simulations for cylinders with flat and hemispherical caps. Resolving the end regions for extreme aspect ratios up to $1/h=10^{100}$ required exponentially stretched surface grids, supplemented by mesh subdivision and singularity-removal techniques (see SM). To stress-test the scalings derived from our intermediate matching, we probe the deepest point within the end region: the very tip of the conductor. As shown in Fig.~\ref{fig:bem}, the numerical results confirm that the surface charge density at that point exhibits the predicted $\mathcal{L}^{-1/2}h^{-1}$ scaling. 

The intermediate behavior \eqref{intermediate final} and its associated square-root-log scaling are not a peculiarity of the present electrostatics problem, but rather a fundamental feature of truncated slender bodies across diverse physical scenarios. To begin with, the problems of steady mass or heat diffusion from a perfectly absorbing or isothermal body, respectively, are mathematically equivalent to the electrostatic problem. Consequently, the surface flux density in these problems is enhanced by $O(\mathcal{L}^{1/2})$ upon approaching the ends. More generally, we anticipate similar behavior whenever slender-body theory yields a unidimensional integral equation governing the density of $1/r$-type sources; this occurs in numerous problems across electrostatics, diffusion, Debye--H\"uckel theory, Stokes flow, elasticity and acoustics \footnote{This is not the case for irrotational flow over a slender body, where the source density is directly prescribed by the local tapering rather than solved for via an integral equation.}. In what follows, we illustrate this universality by considering a modified electrostatics setup, plasmonic resonances, and Stokes flow. 

\textit{External field}.---Consider a thin conductor subjected to an axial external field $E_{\infty}\be_z$. In this scenario, it is convenient to normalize the potential by $E_{\infty}L$ and the linear charge density by $\epsilon E_{\infty}L$. The far-field condition is now $\varphi=-z+o(1)$ as $r^2+z^2\to\infty$, a form which fixes the reference of the potential. By superposition, we can subtract the solution for a uniform surface potential with no external field, to be left with the problem of a zero-potential body in the external field. 

The slender-body formulation is slightly modified. The outer approximation \eqref{outer} now holds for the disturbance potential $\varphi+z$, while the constant term is omitted from the inner approximation \eqref{inner}. Matching thus leads to the integral equation \eqref{integral equation} with the right-hand side replaced by $2\pi z$. The inverse-log expansion \eqref{q expansion} and the recurrence relation \eqref{q recurrence} still apply, but the leading term is now nonuniform, $q_0= 2\pi z$. For a truncated slender body we recover the intermediate behavior \eqref{intermediate final} (with a sign change near $z=-1$). Hence, the end-region scalings remain identical, with field enhancements now measured relative to $E_{\infty}$. 

\textit{Plasmonics}.---Consider a slender metallic nanoparticle irradiated at an angular frequency $\omega$ corresponding to a wavelength $\gg L$. In this quasi-static regime, the near field is irrotational and possesses the spectral representation \citep{Klimov:14}
\begin{equation}\label{spectral}
\mathbf{E} = \sum_n \frac{\epsilon_m(\omega)-1}{\epsilon_m(\omega)-\epsilon\ub{n}}\frac{\langle \mathbf{E}\ub{n},\mathbf{E}\ub{i}\rangle}{\langle \mathbf{E}\ub{n},\mathbf{E}\ub{n}\rangle}\mathbf{E}\ub{n},
\end{equation}
where $\epsilon_m(\omega)$ is the complex  permittivity of the metal, defined here relative to the dielectric background; $\mathbf{E}\ub{i}$ is the incident field; $\epsilon\ub{n}$ and $\mathbf{E}\ub{n}$ are the relative-permittivity eigenvalue and eigenfield of the $n$th localized-surface-plasmon mode (governed by the quasi-static near-field problem in the absence of the incident field); and the inner product denotes a volume integral of the dot product over the interior body domain. The eigenvalues are real and negative, while the eigenfields can be chosen real. The $n$th mode may be resonantly excited near an angular frequency $\omega_r$ for which $\mathrm{Re}[\epsilon_m(\omega_r)]=\epsilon\ub{n}$, provided the absorptive damping is sufficiently weak, $\mathrm{Im}[\epsilon_m(\omega_r)]\ll\mathrm{Re}[\epsilon_m(\omega_r)]$. 

For a slender body of revolution, the dominant modes are longitudinal (axisymmetric), characterized by large-magnitude eigenvalues, and interior eigenpotentials varying primarily along the axis. \citet{Ruiz:19} developed a slender-body theory describing these modes. They found $\epsilon\ub{n}\sim -h^{-2}\mathcal{E}(h)$ \footnote{For a metal described by the Drude model, this large permittivity scaling corresponds to a resonant wavelength scaling as $L/h$, up to logarithmic factors.}, where $\mathcal{E}(h)$ is a reduced eigenvalue governed by the integro-differential system
\begin{subequations}
\label{plasmonic problem}
\begin{gather}
q(z;h)+ \pi\mathcal{E}(h) \frac{d}{dz}\left[\kappa^2(z) \frac{d}{dz}v(z;h)\right] = 0, \\
q(z;h)\ln \frac{2\sqrt{1-z^2}}{h\kappa(z)}+\mathcal{S}[q](z;h)=2\pi v(z;h).
\end{gather}
\end{subequations}
Here $v(z;h)$ and $q(z;h)$ are, respectively, the dimensionless eigenvoltage and linear polarization-charge eigendensity, both of which, along with $\mathcal{E}(h)$, may depend logarithmically upon $h$. Physically, (\ref{plasmonic problem}a) represents Gauss's law applied to a small segment of the body, while (\ref{plasmonic problem}b) generalizes the charge--voltage relation \eqref{integral equation} to account for a varying voltage profile. The appropriate boundary conditions for (\ref{plasmonic problem}a) follow from matching the inner and end-region fields \emph{within} the body. The solutions in \citep{Ruiz:19} focus on needle-like shapes for which  the eigenvoltage satisfies regularity boundary conditions. Here, we consider truncated shapes; in that case, the eigenvoltage satisfies the Neumann boundary conditions $v'(\pm 1;h)=0$, where $v'=dv/dz$. 

We normalize the eigenvoltage to order unity and posit the expansion $v(z;h)\sim \sum_{n=0}^\infty \mathcal{L}^{-n}v_n(z)$. The corresponding expansions for the eigenvalue $\mathcal{E}(h)$ and charge density $q(z;h)$ are then of the form \eqref{q expansion}. At leading logarithmic order, (\ref{plasmonic problem}b) yields $q_0=2\pi v_0$, whereby (\ref{plasmonic problem}a) gives $(\kappa^2 v_0')'+(2/\mathcal{E}_0)v_0=0$. For cylindrical shapes, $\kappa(z)\equiv 1$, the Neumann boundary conditions furnish the eigenvalues as $\mathcal{E}_0=8/(n\pi)^2$ for $n=1,2,\ldots$, with trigonometric eigenfunctions; for the fundamental dipolar mode ($n=1$), $v_0\propto \sin(\pi z/2)$. With both $v(z;h)$ and $\kappa(z)$ bounded and non-vanishing at the ends, (\ref{plasmonic problem}b) dictates that $q(z;h)$ develops the intermediate behavior \eqref{intermediate final}, up to normalization-dependent prefactors. Accordingly, the electric eigenfield in the end regions scales as  $O(\mathcal{L}^{-1/2}h^{-1})$. 

For a longitudinally polarized incident plane wave, substituting this eigenfield into the spectral representation \eqref{spectral} yields an estimate for the resonant end-region field. Because the interior eigenfield is dominated by longitudinal voltage variations and $\epsilon_m(\omega)\sim -h^{-2}\mathcal{E}(h)$ near resonance, the end-region field is enhanced by $O[h^{-3}\mathcal{L}^{-1/2}/\mathrm{Im}(\epsilon_m(\omega_r))]$ relative to the incident field. While this is weaker than the $O[h^{-4}\mathcal{L}^{-1}/\mathrm{Im}(\epsilon_m(\omega_r))]$ enhancement for a needle-like nanoparticle, the cylindrical end-region field may be locally amplified near a sharp rim as discussed for the electrostatic problem. These resonant ``hotspots'' are key to  nanoplasmonics applications \citep{Maier:07}, from amplifying nonlinear light-matter interactions and surface-enhanced Raman scattering to targeted photothermal destruction of tumors. 

\textit{Stokes flow}.---Consider a slender body of revolution translating along its axis with velocity $U\be_z$ in an otherwise quiescent viscous fluid. At zero Reynolds number, the flow is governed by the Stokes equations, for which slender-body theory is highly developed \citep{Batchelor:70,Cox:70,Tillett:70,Keller:76,Johnson:80}; nonetheless, as with electrostatics, end effects for truncated shapes have not been resolved. The longitudinal-scale linear force density $f(z;h)$ exerted on the fluid, normalized by $\mu U$ (where $\mu$ is the dynamic viscosity), satisfies
an integral equation like \eqref{integral equation} but for the replacement of $\kappa(z)$ by $\tilde{\kappa}(z)=e^{1/2}\kappa(z)$. Consequently, for truncated shapes, the intermediate behavior \eqref{intermediate final} holds and implies $O(\mathcal{L}^{-1/2}h^{-1})$ stresses in the end regions \footnote{The transverse translation problem reduces to a similar integral equation, leading to the same stress scaling.}.  

As with the electrostatic capacitance, the end regions do not contribute to the net drag at leading algebraic order, given by $\mu U L\int_{-1}^1f(z)\,dz$. In other Stokes-flow problems, however, they may have global ramifications. Indeed, consider a solid ``active'' particle self-propelling in the absence of external loads via a tangential slip-velocity distribution prescribed on its boundary. Assuming axial symmetry, the Lorentz reciprocal theorem yields the swimming speed as an integral over the particle boundary \citep{Stone:96},
\begin{equation}\label{reciprocal}
\mathcal{U}=\frac{1}{\mathcal{D}^*}\oint (\bn\bcdot \bm{\sigma}^*) \bcdot \bu_s\,dA,
\end{equation}
where $\bn$ is the unit normal pointing into the fluid, $\bu_s$ is the slip distribution, and $\bm{\sigma}^*$ and $\mathcal{D}^*$ are, respectively, the stress tensor and net drag in the passive translation problem; here velocities are normalized by a characteristic slip velocity $W$, stresses by $\mu W/L$ and drag by $\mu W L$. For truncated slender bodies, $\bm{\sigma}^*$ is of the same \textit{algebraic} order over the $O(1)$ body scale and within the $O(h)$ end regions. Thus, if the slip velocity $\bu_s$ is larger in the end regions by an $O(h^{-1})$ algebraic factor, the end regions may contribute to the surface integral in \eqref{reciprocal}, and thus to the swimming speed, at leading algebraic order. Such a scenario naturally occurs for rod-shaped phoretic Janus particles, where the slip is driven by the surface gradient of solute molecules emitted or absorbed at the surface. As discussed by \citet{Yariv:19:Langmuir}, systematically addressing this scenario remains an open problem. Indeed, past analyses \citep{Golestanian:07,Schnitzer:15,Ibrahim:18,Katsamba:20,Katsamba:22} have overlooked the anomalous end structure uncovered herein.
 
\textit{Discussion}.---We have obtained the scaling of the charge density near the ends of thin conducting cylinders, a longstanding electrostatics problem going back to Maxwell; closed a fundamental gap in asymptotic slender-body theory, where the end regions of truncated shapes have to date remained unresolved; and demonstrated the universality of the uncovered end structure across broad classes of geometries and physical domains. This theory may inform the modeling of phenomena governed by localized field enhancements and stress concentrations, such as dielectric breakdown, field emission, targeted heating, cavitation and material fracture. Furthermore, we anticipate that in certain scenarios, as in the aforementioned active-particle problem, this structure may also dictate global outputs. 

\bibliography{refs}

\end{document}